\documentclass[epj]{svjour}
\usepackage{graphics}
\begin{document}
\title{Inclusive pentaquark and strange baryons
production\\ in hadron beam experiments  at high energy}
\author{I.M.Narodetskii
\and M.A.Trusov
\and A.I.Veselov
}                     
\institute{ITEP, Moscow, Russia
}
\date{Dec. 24, 2006}
%
\abstract{We estimate the high-energy behavior of the $\Theta^+$
and $\Lambda(1520)$ production cross sections in inclusive $pp$
collisions  using the $K$ exchange diagram. We show that the cross
section of the $\Theta^+$-production is suppressed compared to the
production of $\Lambda(1520)$. As a byproduct we also estimate the
contribution of the $\pi$ exchange diagram for the inclusive
$\Lambda(1520)$ production in $\Sigma p$ collisions.
%
} 
\authorrunning{I.~M.~Narodetskii~{\it et al}~}
\titlerunning{Inclusive pentaquark and strange baryons production}
\maketitle
\section{Introduction}

The possible existence of the $\Theta^+$ pentaquark remains one of
the puzzling mysteries  of recent years. To date there are more
than 20 experiments with evidence for this state, but criticism
for the $\Theta^+$ claim arises because similar number of high
energy experiments did not find any evidence for the $\Theta^+$,
even though the other  ``conventional'' three-quark hyperons  such
as $\Lambda(1520)$ hyperon resonance are seen clearly
\cite{qnp06}.


Most of negative high energy experiments are high statistic hadron
beam experiments. {\it E.g.} HERA-B, a fixed target experiment at
the 920 GeV proton storage ring of DESY \cite{HERA-B} finds no
evidence for narrow signals in the $\bar K^0_Sp$ channel and only
sets modest upper limits for $\Theta^+$ production of less than
16~$\mu$b/N and less than about 12$\%$ relative to $\Lambda(1520)$
in mid-rapidity region. This negative result would present serious
rebuttal evidence to worry about. However, without obvious
production mechanism of the $\Theta^+$ (if it exists) or even
$\Lambda(1520)$ the rebuttal is not very convincing.

In this paper we estimate the high-energy behavior of the
$\Theta^+$ and $\Lambda(1520)$ production cross sections in
inclusive $pp$ collisions  using the $K$ exchange diagram, which
is known to survive at high energies in the beam/target
fragmentation region. We show that the cross section of the
$\Theta^+$-production is suppressed compared to the production of
$\Lambda(1520)$. This suppression is mainly due to the smallness
of the coupling constant $G^2_{\Theta KN}$ compared  to
$G^2_{\Lambda KN}$ that in turn is related to the small width of
the $\Theta^+$. As a byproduct we also estimate the contribution
of the $\pi$ exchange diagram for the inclusive $\Lambda(1520)$
production in $\Sigma p$ collisions.

\section{Inclusive cross sections}

\begin{figure}
\begin{center}
\resizebox{1.00\width}{!}{
\includegraphics{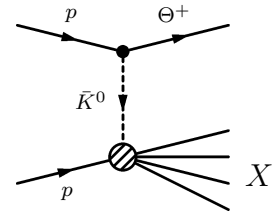}}
\vspace{0.2cm} \caption{\label{fig:pp_theta} The ${\bar K}^0$
exchange diagram for the $\Theta^+(1540)$ production in inclusive
pp scattering
}
\end{center}
\end{figure}

We assume that the $\Theta^+$ exists and
$J^P(\Theta^+)={\frac{1}{2}}^+$. Consider the $\Theta^+$
production in the reaction $ p+p\to\Theta^++X$,
where $X$ is
unspecified inclusive final state carrying the strangeness -1. The
${\bar K^0}$ exchange diagram for $pp\to\Theta^+X$ is shown in
Fig. \ref{fig:pp_theta}.

The standard expression for the contribution of this diagram
written
in terms of the 4-momentum transfer squared $t=q^2$ and the
invariant mass $W$ of the $\bar K^0p$ system is well known
\cite{Yao}. At high energy, it is more convenient to convert this
expression to an integral over the Feynman variable $x_F$, the
fraction of the incident proton momentum carried by the $\Theta^+$
in the initial direction of the proton (in the center-of-mass
system), and $k_{\bot}$, the transverse momentum of $\Theta^+$
relative to the initial proton direction. Then the contribution of
the $K$ meson exchange to the double differential cross section
for the $\Theta^+$ inclusive production reads
\begin{equation}
\label{dsigma}
\frac{d\sigma}{dx_Fdk_{\bot}^2}=\frac{1}{4\pi}\,\frac{G_{\Theta
KN}^2}{4\pi}\, J\, \frac{p}{E_{\Theta}}\,
\Phi_{p\,\Theta}(t)\,F^4(t)\, \sigma_{\rm {tot}}^{\bar K^0
p}(s_1),
\end{equation}
where $G_{\Theta KN}$ is the coupling constant for the decay
$\Theta^+\to {\bar K}^0p$,
$E_{\Theta}=\sqrt{x_F^2p^2+k_{\bot}^2+m_{\Theta}^2}$ is the
$\Theta^+$ energy in the center-of-mass system,
$s_1=W^2=s+m_{\Theta}^2-2E_{\Theta}\sqrt{s}$,
${s}=4(p^2+m_p^2)$ is the center-of-mass energy squared, $p$ is
the center-of-mass momentum, and
$t=m_p^2+m_{\Theta}^2-2E_{\Theta}\sqrt{p^2+m_p^2}+2x_Fp^2$ .
The factor $J$ is the ratio of flux factors in the $pp$ and $\bar
K^0p$ reactions. To evaluate the cross sections away from the pole
position $t=M_{K}^2$ we include  the phenomenological form factor
$F_{K}(t)$. The function $\Phi_{p\,\Theta}(t)$ is the squared
product of the vertex function for $p\to\Theta^+\bar K^0$ and the
kaon propagator:
\begin{equation}
\label{phi}
\Phi_{p\,\Theta}(t)=\frac{(m_p-m_{\Theta})^2-t}{(t-m_K^2)^2}.
\end{equation}

\noindent In the high energy limit with accuracy
$\mathcal{O}(1/p^2)$
\begin{equation}\label{hel}
J\cdot \frac{p}{E_{\Theta}}\approx\frac{1-x_F}{x_F},\end{equation}
\begin{equation}\label{hell}
s_1\approx (1-x_F)s,~~~ t\approx
m_{\Theta}^2+m_p^2(1-x_F)-\frac{m_{\Theta}^2+k_{\bot}^2}{x_F},
\end{equation}
and   the double differential cross section written in terms
$x_F$ and $k_{\bot}^2$ reads
\begin{equation}
\label{dsigma_he} \frac{d\sigma}{dx_Fdk_{\bot}^2}
=\frac{1}{4\pi}\frac{G_{\Theta KN}^2}{4\pi}\cdot \frac{1-x_F}{x_F}
\Phi_{p\,\Theta}(t)F^4(t)\sigma_{\rm{tot}}^{{\bar K}^0 p}(s_1).
\end{equation}

\subsection{The $\Theta^+KN$ vertex}

\noindent The $\Theta^+ KN$ vertex
 is
\begin{equation}
\label{NKTvertex} { L}_{\Theta KN}=iG_{\Theta
KN}(K^{\dag}{\bar\Theta}\gamma_5N+\bar N\gamma_5\Theta K),
\end{equation} with the operator $\gamma_5$ corresponding to
positive $\Theta^+$ parity. The Lagrangian (\ref{NKTvertex})
corresponds with the $\Theta^+$ being a $p$-wave resonance in the
$K^0p$ system. The partial decay width $\Gamma_{\Theta\to  K^0p}$
is
\begin{equation}\label{NKTwidth}\Gamma_{\Theta\to K^0p}=\frac{G^2_{\Theta
KN}}{4\pi}\cdot\frac{2p_K^3}{(m_{\Theta}+m_p)^2-m_K^2} ,
\end{equation} where $p_K=260$ MeV/c is the kaon momentum in the
rest frame of $\Theta^+$. To extract the value for $G_{\Theta
KN}$, we need the experimental information of the width
$\Gamma_{\Theta K N}$, which is not known precisely but whose
measurement is the subject of several planned dedicated
experiments. To provide numerical estimates, we will use the value
$\Gamma_{\Theta \to K^0 p}=1$~MeV. This corresponds to the full
width $\Gamma_{\Theta KN}=\Gamma_{\Theta\to K^0
p}+\Gamma_{\Theta\to K^+ n}=2$~MeV, which is consistent with the
upper limit for the width derived from elastic $KN$
scattering~\cite{Cahn:2003wq}~\footnote{ An additional reason for
the smallness of the pentaquark width arises in the string model,
in which  the pentaquark decay is accompanied by the annihilation
of the two string junctions. Indeed, the pentaquark containing
three string junctions dissociates ``fall apart'' into two minimal
color singlets containing only one string junction. The
annihilation of the string junctions may produce a complimentary
smallness of the width.}.

Evaluating Eq.~(\ref{NKTwidth}) with $\Gamma_{\Theta \to
K^0p}=1$~MeV, we extract the value $G_{\Theta KN}$
\begin{equation}
\frac{G_{\Theta KN}^2}{4\pi}=0.167\cdot\frac{ \Gamma_{\Theta\to
K^0p}}{1~\rm{MeV}}, \end{equation}
 which will be used in
the subsequent estimates for the inclusive cross section.

\subsection{$\Lambda(1520)\,KN$ vertex}

\noindent The $\Lambda(1520)\,KN$ vertex is
\begin{equation}
\label{KNL}
 L_{\Lambda KN}=\frac{G_{\Lambda KN}}{m_K}\left(\bar
\Lambda^{\mu}\gamma_5N\partial_{\mu}K+\bar
N\gamma_5\Lambda^{\mu}\partial_{\mu}K^{\dag}\right),
\end{equation} where $\Lambda^{\mu}$ is the vector spinor for
the spin 3/2 particle. The Lagrangian  (\ref{KNL}) corresponds
with the $\Lambda(1520)$ being a $d$-wave resonance in the $K^-p$
system. The  $\Lambda(1520)\to pK^-$ width is
\begin{equation}\Gamma_{\Lambda \to K^-p}=\frac{G_{\Lambda KN}^2}{4\pi}\cdot
\frac{2p_K^5}{3m_K^2}\cdot\frac{1}{(m_{\Lambda}+m_p)^2-m_K^2},
 \end{equation} where $p_K=246~\mathrm{MeV/c}$ is the
kaon momentum in the rest frame of $\Lambda(1520)$. Using the PDG
values of $\Gamma_{\rm{tot}}(\Lambda(1520))=15.6$ MeV and
$\rm{Br}(\Lambda(1520)\to N\bar K)=45\%$ we obtain
\begin{equation}
\frac{G_{\Lambda  KN}^2}{4\pi}\approx 8.14.
\end{equation}
The function $\Phi_{p\,\Lambda}(t)$
is
\begin{equation}\label{eq:phi-p-lambda} \Phi_{p\,\Lambda}(t)
=\frac{(m_p+m_{\Lambda })^2-t}{6m_{\Lambda }^2m_K^2}\cdot
\frac{((m_p-m_{\Lambda })^2-t)^2}{(t-m_K^2)^2}.\end{equation} The
expression (\ref{eq:phi-p-lambda}) includes the factor $1/m_K$ in
(\ref{KNL}).

\section{Results}


The total cross section for the general case of the fragmentation
of a baryon $a$ into a baryon $b$ due to the exchange by the meson
$m$ is
\begin{equation}
\label{eq:cs_total} \sigma_{ab}\,=\,\frac{G^2_{bma}}{4\pi}\,\int
dx_F\int
dk^2_{\bot}\,K_{ab}(x_F,k^2_{\bot})\,\sigma_{\rm{tot}}^{mp}(s_1),
\end{equation} where
$K_{ab}(x_F,k^2_{\bot})\,=\,(1-x_F)\,\Phi_{ab}(t)\,F^4(t)/x_F$. We
employ two representative examples for the form factor $F(t)$:

%
$$
\mathrm{A:}~~F(t)=\frac{\Lambda^2-m_K^2}{\Lambda^2-t},~~~~~
\mathrm{B:}~~F(t)=\frac{\Lambda^4}{\Lambda^4+(t-m_K^2)^2},
$$
the cut-off parameter $\Lambda$ being a typical hadronic scale
$\Lambda=1$~GeV.

Because of (\ref{hel}), (\ref{hell}) all the energy dependence of
the right hand side of Eq. (\ref{dsigma_he}) is due to the factor
$\sigma^{mp}(s_1)$. Since $\sigma^{mp}(s_1)$ is slow varying
function of $s_1=(1-x_F)s$ everywhere, except the low energy
region, we can take it out of the integral at the point $\hat
s_1=(1-\hat x_F)s$, where $\hat x_F$ is the point at which
$d\sigma^{mp}/dx_F$ reaches the maximum\footnote{The typical
values of $\hat x_F$ are $\sim 0.8-0.9$, depending on the reaction
considered. Therefore, in the fragmentation region, the effective
energy $\sqrt{\hat s_1}$  {\it is always much smaller} than
$\sqrt{s}$, but at high energies {\it is still large enough} to
use the asymptotic of $\sigma^{mp}$}. Then we obtain
\begin{equation} \sigma_{ab} \approx \frac{G^2_{bma}}{4\pi} \sigma_{\rm{tot}}^{mp}(\hat
s_1){\hat K}_{ab},\end{equation} where the quantities
\begin{equation}{\hat K}_{ab}\,=\,\int d{x_F}\int dk^2_{\bot}\,K_{ab}(x_F,k^2_{\bot})\end{equation}
 do not depend on energy, and $\sigma_{\rm{tot}}^{mp}(\hat
s_1)$ is a constant up to logarithmic and power corrections.

 For estimation we take the total cross sections
$\sigma_{\rm{tot}}^{\bar K^0p}$ and $\sigma_{\rm{tot}}^{K^+p}$ to
be a constant ($\sigma_{\rm{tot}}^{\bar
K^0p}\sim\sigma_{\rm{tot}}^{K^+p}\sim 20~\rm{mb}$ at
$\sqrt{s_1}~>~ 10$~GeV.)  Then we obtain for the production cross
sections\footnote{The values in (\ref{numerics}) correspond to the
region $x_F ~>~ 0$. For $pp$ scattering the total cross sections
are two times larger.}

\begin{equation}\label{numerics}
\sigma(pp\to \Theta^+(1540)X)=
0.8\,(1.6)\times\frac{\Gamma_{\Theta\to
K^0p}}{1~\rm{MeV}}~\mu\mathrm{b},\end{equation}\begin{equation}
\sigma(pp\to \Lambda^+(1520)X)=106\,(126)~\mu\mathrm{b},
\end{equation} where the first values refer to the form factor (A) and
the second ones to the form factor (B).
The result for $\sigma(pp\to \Theta^+X)$ matches well that of Ref.
\cite{Vera} for the inclusive $pp\to\Theta^+X$ production
 at $\sqrt{s}~<~ 10$~GeV.
If $\Gamma_{\Theta KN}=0.36\pm 0.11$~MeV as is claimed in
\cite{diana}, our result for the $\Theta^+$ production cross
section should be correspondingly smaller.

\begin{figure}
\begin{center}
\resizebox{0.30\width}{!}{
\includegraphics{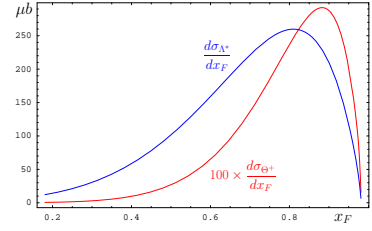}}
\caption{\label{fig:pp_inclusive:x} $x_F$ dependence of the
inclusive $pp\to \Theta^+(1540)$ and $\Lambda(1520)$ cross
sections
}
\end{center}
\end{figure}

In scattering hadronic probes at high energy from nuclear target
the only positive signal for the $\Theta^+$ decaying to $K_S^0p$
 was reported by the SVD Collaboration, using 70 GeV proton in
 a fixed target arrangement $pA\to \Theta^+X$ at a center-of-mass energy of
 about 11.5 GeV \cite{SVD-1}.
 Our prediction for $\sigma(pp\to \Lambda(1520)^+X)$ agrees with
the preliminary result of the SVD-2 collaboration, but
$\sigma(pp\to \Theta^+X)$ is lower than the preliminary cross
section estimation (for $x_F>0$) : $\sigma\cdot
\rm{Br}(\Theta^+\to pK^0)\sim 6~\mu$b. The illustrative examples
of the $x_F$  distributions for the $\Theta^+$ and $\Lambda(1520)$
are shown in Fig. \ref{fig:pp_inclusive:x}.  for the form factor
A.

The ratio of $\Theta^+$ to $\Lambda(1520)$ production
cross-sections is $\sim 1\%$. Our estimation is a bit larger than
that obtained in the fragmentation-recombination model
 \cite{titov} but still is rather small and probably can be useful to explain why the
$\Theta^+$ production is suppressed in some high energy
experiments.

In the same way it is possible to make quantitative predictions
for other type of colliding particles. As an example, we estimate
the cross section for the inclusive $\Lambda(1520)$ production in
$\Sigma^- p\to\Lambda(1520)$ collisions  at 600 GeV/c studied in
the fixed target Fermilab experiment E781 (SELEX). In the
fragmentation region of the $\Sigma$-hyperon this reaction can
proceed via the $\pi$-meson exchange.
Using
\begin{equation}
\Gamma_{\Lambda\to \pi^-\Sigma^+}=\frac{1}{3}\cdot
\rm{Br}(\Lambda\to
\pi\Sigma)\cdot\Gamma_{\rm{tot}}\\{}=2.18~\rm{MeV},\end{equation}
$G_{\Lambda\pi\Sigma}^2/4\pi\approx 0.353$,
 and $\sigma(\pi N)=25~\mu$b we get
\begin{equation}
\sigma(\Sigma p\to \Lambda(1520)X)= 314\,(340)~\mu\mathrm{b},
\end{equation}
where, as before, the first value refer to the form factor (A) and
the second ones to the form factor (B). For the ratio of inclusive
$\Sigma p$ and $pp$ cross sections we get
\begin{equation}
\frac{\sigma(\Sigma p\to\Lambda(1520)X)}{\sigma(p
p\to\Lambda(1520)X)}\approx 2.9\,(2.7),
\end{equation}
that agrees with the preliminary experimental result ($\approx
2.6$)
of the SELEX collaboration \cite{SELEX}.

\section{Conclusions}

Let us recall that our estimations may somehow depend on specific
assumptions regarding for instance the $K$-meson exchange
dominance at forward direction, and on the choice of the form
factor. As an outlook, it would be interesting to go beyond the
present calculation and to perform a systematic study of $K$,
$K^*$ and $\pi$ Regge exchanges into inclusive production of
(anti)strange baryons in $pp$ collisions. We plan to come back to
these issues in a next publication.

This work was supported by RFBR grants 04-02-17263, 05-02-17869,
06-02-17120, and by the State Contract\\ 02.445.11.7424, part
2006-RI-112.0/001/398.

\end{document}